\documentclass[aps,prl,reprint,superscriptaddress]{revtex4-1}  

\usepackage{amsmath}
\usepackage{empheq}
\usepackage{braket}
\usepackage{amssymb}
\usepackage{graphicx}
\usepackage{color}
\usepackage{isotope}
\usepackage{rotating} 
\usepackage{siunitx}
\usepackage{hyperref}
\usepackage{times,txfonts}
\usepackage{wasysym}
\usepackage{tikz}

\newcommand{\refstateket}{\ket{\Psi_{\text{ref}}}}
\newcommand{\refstatebra}{\bra{\Psi_{\text{ref}}}}

\newcommand{\Nmaxref}{N^{\text{ref}}_{\text{max}}}

\newcommand{\hbarOmega}{\hbar\Omega} 

\newcommand{\elem}[2]{\isotope[#2]{#1}}

\newcommand{\HamNOzeroSymbol}{E}

\newcommand{\MEzeroHamNO}{\HamNOzeroSymbol}

\newcommand{\Nmax}{\ensuremath{N_{\text{max}}}} 

\newcommand{\emax}{e_{\text{max}}} 
\newcommand{\IMSRGpara}{s}

\newcommand{\SRGpara}{\alpha}
\newcommand{\ddIMSRGpara}[1]{\frac{ \mathrm{d} {#1} }{ \mathrm{d}\IMSRGpara} } 
    
\newcommand{\comm}[2]{ \big[ {#1}, {#2} \big] }          

\newcommand{\symboldiamond}[1][black]{{\color{#1}\hspace{-1pt}\footnotesize\begin{turn}{45} $\blacksquare$ \end{turn}}}

\newcommand{\symboltriangleup}[1][black]{{\color{#1}{\raisebox{1.5ex}{\begin{turn}{180}$\blacktriangledown$\end{turn}}}}}
\newcommand{\symboltriangledown}[1][black]{{\color{#1}$\blacktriangledown$}}

\newcommand{\symboltriangleright}[1][black]{{\color{#1}{\begin{turn}{90}$\blacktriangledown$\end{turn}}}}
\newcommand{\symbolbox}[1][black]{{\color{#1}$\blacksquare$}}
\newcommand{\symbolcircle}[1][black]{{\color{#1}\large$\bullet$}}

\newcommand{\symbolbigstar}[1]{{\color{#1}$\bigstar$}}

\definecolor{FGViolet}{rgb}{0.61,0.32,0.61}
\definecolor{FGDarkBlue}{rgb}{0,0,0.6}
\definecolor{FGBlue}{rgb}{0,0,0.8}
\definecolor{FGLightBlue}{rgb}{0.2, 0.6, 0.8}
\definecolor{FGGreen}{rgb}{0.2,0.7,0.2}
\definecolor{FGLightGreen}{rgb}{0.4,1,0.4}
\definecolor{FGYellow}{rgb}{1,0.95,0}
\definecolor{FGOrange}{rgb}{0.95,0.5,0.1}
\definecolor{FGRed}{rgb}{0.8,0,0}
\definecolor{FGWhite}{rgb}{1,1,1}
\definecolor{FGLightGray}{rgb}{0.8,0.8,0.8}
\definecolor{FGGray}{rgb}{0.5,0.5,0.5}
\definecolor{FGDarkGray}{rgb}{0.3,0.3,0.3}
\definecolor{FGBlack}{rgb}{0,0,0}

\definecolor{myblue}{rgb}{0, 0, 1}
\definecolor{myred}{rgb}{1.0, 0, 0}
\definecolor{mygreen}{rgb}{0.0, 0.5, 0.0} 
\definecolor{darkviolet}{rgb}{0.58039, 0.0, 0.82745}
\definecolor{mycyan}{rgb}{0.0, 0.75, 0.75}
\definecolor{myorange}{rgb}{1.0, 0.64705, 0.0}
\definecolor{mybrown}{rgb}{0.64705, 0.164705, 0.164705}

\newcommand{\NmaxZeroSymbol}{\symbolcircle[myblue]}
\newcommand{\NmaxTwoSymbol}{\symbolbox[myred]}
\newcommand{\NmaxFourSymbol}{\symboltriangleup[mygreen]}
\newcommand{\NmaxSixSymbol}{\symboldiamond[darkviolet]}
\newcommand{\NmaxEightSymbol}{\symbolbigstar{mycyan}}
\newcommand{\NmaxTenSymbol}{\symboltriangledown[myorange]}
\newcommand{\NmaxTwelveSymbol}{\symboltriangleright[mybrown]}

\newcommand{\IMNCSMsymbol}{\symbolbox[myred]}
\newcommand{\MRIMSRGHFBsymbol}{\symboltriangleup[mygreen]}
\newcommand{\ITsymbol}{\symbolcircle[myblue]}

\begin{document}

\title{
\textit{Ab Initio} Description of Open-Shell Nuclei:\\
Merging No-Core Shell Model and In-Medium Similarity Renormalization Group
}

\author{Eskendr Gebrerufael}
\email[]{eskendr.gebrerufael@physik.tu-darmstadt.de}
\affiliation{Institut f\"ur Kernphysik, Technische Universit\"at Darmstadt, Schlossgartenstr.\ 2, 64289 Darmstadt, Germany}

\author{Klaus Vobig}
\affiliation{Institut f\"ur Kernphysik, Technische Universit\"at Darmstadt, Schlossgartenstr.\ 2, 64289 Darmstadt, Germany}

\author{Heiko Hergert}
\affiliation{NSCL/FRIB Laboratory and Department of Physics \& Astronomy, Michigan State University, East Lansing, MI 48824-1321, USA}

\author{Robert Roth}
\affiliation{Institut f\"ur Kernphysik, Technische Universit\"at Darmstadt, Schlossgartenstr.\ 2, 64289 Darmstadt, Germany}

\date{\today}

\begin{abstract}

We merge two successful \emph{ab initio} nuclear-structure methods, the no-core shell model (NCSM) and the multi-reference in-medium similarity renormalization group (IM-SRG) to define a new many-body approach for the comprehensive description of ground and excited states of closed and open-shell nuclei.
Building on the key advantages of the two methods---the decoupling of excitations at the many-body level in the IM-SRG and the access to arbitrary nuclei, eigenstates, and observables in the NCSM---their combination enables fully converged no-core calculations for an unprecedented range of nuclei and observables at moderate computational cost.
We present applications in the carbon and oxygen isotopic chains, where conventional NCSM calculations are still feasible and provide an important benchmark.
The efficiency and rapid convergence of the new approach make it ideally suited for \emph{ab initio} studies of the complete spectroscopy of nuclei up into the medium-mass regime.

\end{abstract}


\maketitle

\paragraph{Introduction.}
One of the most dynamic areas in nuclear structure theory today is the development of \emph{ab initio} many-body methods for the comprehensive description of open-shell nuclei.
This includes not only ground states, but also low-lying excitations and spectroscopic observables.
Traditionally, nuclear spectroscopy is the domain of shell-model-type approaches, both the valence-space shell model \cite{RevModPhys.77.427} and the \emph{ab initio} no-core shell model (NCSM) \cite{NaGuNo07,BROWN2001517,Barrett2013131}.
These methods solve a large-scale Hamiltonian eigenvalue problem in a truncated model space and address ground and excited states on equal footing, but they are limited by the basis dimension \cite{Vary2009}.

Several other methods have been developed that tackle the many-body problem from a different angle, among them the coupled-cluster (CC) approach \cite{Binder2014119,PhysRevC.87.021303,Coester1958421,PhysRevLett.94.212501,PhysRevLett.101.092502,HaPa07-CC} and the in-medium similarity renormalization group (IM-SRG) \cite{PhysRevLett.106.222502,PhysRevC.85.061304,Hergert2016165,Hergert:2016etg}.
Instead of solving the eigenvalue problem directly, these methods use a similarity transformation to decouple a reference state, representing the ground state, from all particle-hole excitations.
This concept of decoupling is very powerful and complementary to a direct NCSM-type diagonalization.
Generally, CC and IM-SRG have different computational characteristics and a much better scaling with particle number, but their basic formulation is limited to ground states.
Their complementarity suggests that a combination of both philosophies, direct diagonalization and many-body decoupling, could be advantageous.
First steps along these lines are the effective interactions for the valence-space shell-model extracted from CC and IM-SRG calculations presented recently \cite{PhysRevLett.113.142501,PhysRevC.93.051301,PhysRevLett.113.142502,Stroberg:2016ung,PhysRevC.94.011301}.

In this Letter, we merge NCSM and IM-SRG into a new \emph{ab initio} many-body tool to universally address ground and excited states of closed and open-shell nuclei up to medium masses.
After discussing the method with its advantages and limitations we benchmark it against direct NCSM calculations for several carbon and oxygen isotopes.
In contrast to valence-space methods, we propose an \emph{ab initio} no-core approach, where convergence with respect to all model-space truncations is demonstrated explicitly.

\paragraph{No-Core Shell Model.}

The NCSM is one of the most powerful and universal \emph{ab initio} methods \cite{Barrett2013131,0954-3899-36-8-083101}.
It is built on a representation of the Schr\"odinger equation as a large-scale matrix eigenvalue problem, using an expansion of the eigenstates in an orthonormal basis of $A$-body states, $\ket{\Psi_n} = \sum_{\nu} C_{\nu}^{(n)} \ket{\Phi_{\nu}}$.
Typically, the basis $\ket{\Phi_{\nu}}$ is composed of Slater determinants built from harmonic-oscillator (HO) single-particle states, but other single-particle bases can be used as well \cite{1742-6596-403-1-012014}.
The many-body basis must be truncated to render the problem numerically tractable.
For the NCSM, we typically use the number of HO excitation quanta above the lowest-energy basis states as a truncation parameter $\Nmax$.
Eventually, the finite matrix eigenvalue problem is solved for a few low-lying eigenstates via Lanczos-type algorithms, yielding energies and eigenvectors that can be used to compute any secondary observable.

The truncation of the many-body space constitutes the only departure from an exact treatment of the Schr\"odinger equation and we have to demonstrate that the truncation does not affect the observables of interest.
This convergence is the critical aspect in determining the uncertainties of the method and the limiting factor for its applications.
Several tools are being used to extend reach of the NCSM, e.g., through additional truncations of the many-body model space \cite{PhysRevLett.99.092501,PhysRevC.79.064324} or through a pre-diagonalization of the Hamiltonian by a unitary transformation \cite{PhysRevLett.107.072501,PhysRevLett.103.082501,Okubo01111954,Suzuki01121980}.
Despite these improvements, the NCSM is typically limited to $p$-shell and lower $sd$-shell nuclei \cite{PhysRevC.90.024325,PhysRevC.83.034301}.

\paragraph{In-Medium Similarity Renormalization Group.}

The IM-SRG aims to decouple a pre-defined reference state $\refstateket$ from all excitations via a unitary transformation that is implemented via a flow equation \cite{PhysRevLett.106.222502,PhysRevC.85.061304,PhysRevC.87.034307,PhysRevLett.110.242501,PhysRevC.90.041302,Hergert2016165,Hergert:2016etg}.
The technical tool to make this decoupling in $A$-body space tractable is the normal ordering of all operators with respect to $\refstateket$.
Here, we use on the generalized normal ordering and Wick's theorem of Kutzelnigg and Mukherjee \cite{kutzelnigg:432,Mukherjee1997561,kong:234107}, which can be applied with single and multi-determinantal reference states $\refstateket$.

The Hamiltonian $H(\IMSRGpara)$ and the generator $\eta(\IMSRGpara)$ are normal ordered and inserted into 
$	\ddIMSRGpara{} H(\IMSRGpara)~=~\comm{ \eta(\IMSRGpara) }{ H(\IMSRGpara)  } $
to obtain the multi-reference IM-SRG flow equations \cite{PhysRevLett.110.242501,Hergert:2016etg}.
The generator is designed to suppress pieces of the Hamiltonian that couple the reference state to excited basis determinants orthogonal to the reference state.
We use the so-called imaginary time generator as a compromise between efficiency and robustness (see, e.g., \cite{Hergert2016165}).
To render the flow equations tractable, residual normal-ordered three-body contributions of the initial Hamiltonian and normal-ordered three and multi-particle contributions in the IM-SRG flow equations are discarded.
Thus, we obtain a system of coupled first-order differential equations for the zero, one, and two-body part of the normal-ordered Hamiltonian \cite{PhysRevLett.110.242501,Hergert:2016etg}.
The zero-body part directly yields the expectation value of the Hamiltonian w.r.t.\ the reference state, $E(\IMSRGpara) = \refstatebra  H(\IMSRGpara)\refstateket$, and represents the ground-state energy for $\IMSRGpara \rightarrow \infty$ in standard applications of the IM-SRG.

\paragraph{Merging NCSM and IM-SRG.}

Building on and extending the ideas of both methods, we propose a combination of NCSM and multi-reference IM-SRG for a comprehensive \emph{ab initio} description of open-shell nuclei.
In a first step we perform an NCSM calculation to obtain a reference state for the specific nucleus of interest.
We use the full NN+3N Hamiltonian and solve the NCSM eigenvalue problem in a small model space with truncation parameter $\Nmaxref$.
Here we choose $\Nmaxref=0$ model spaces which are multidimensional in open-shell nuclei---for oxygen isotopes between \elem{O}{16} and \elem{O}{24}, for instance, they contain all neutron configurations in the $sd$ shell. 
The lowest eigenstate with the appropriate quantum numbers serves as reference state $\refstateket$.
In a second step we normal order the Hamiltonian with respect to this multi-determinantal reference state and solve the multi-reference IM-SRG flow equations.
For each value of the flow parameter $\IMSRGpara$, the flow equations yield a normal-ordered Hamiltonian $H(\IMSRGpara)$. Thus, we generate a family of Hamiltonians in which multi-particle-multi-hole excitations are successively decoupled from $\refstateket$.
In the third step, the IM-SRG-evolved Hamiltonians are used in NCSM calculations for a range of truncation parameters $\Nmax$.
These calculations provide ground- and excited-state energies, as well as eigenvectors in a no-core model space.
The eigenvectors can be used to evaluate other observables that have been transformed in a consistent IM-SRG evolution.
For simplicity, we refer to this whole scheme as IM-NCSM, because the key point is the use of an in-medium decoupled Hamiltonian in the NCSM.

The IM-NCSM scheme has a number of important advantages over simple IM-SRG or NCSM calculations.
The initial NCSM calculation can be performed for arbitrary open-shell systems and we can control the complexity of $\refstateket$ via $\Nmaxref$.
The decoupling dramatically accelerates the convergence of the subsequent NCSM calculation, and we obtain ground and excited-state wave functions that can be used for further investigations.
Since we have a continuous mapping between the initial and the decoupled Hamiltonian, we can probe and quantify the effects of truncations of the IM-SRG flow equations by varying the flow parameter.

\paragraph{Calculation Details.}

Calculations in this work use the chiral N$^3$LO NN interaction by Entem and Machleidt with cutoff $\Lambda_{\text{NN}}=\SI{500}{\MeV/c}$ \cite{EnMa2003PhysRevC.68.041001} and local N$^2$LO 3N interaction with $\Lambda_{\text{3N}}=\SI{400}{\MeV/c}$ \cite{Navratil:2007zn,PhysRevC.90.024325}.
This Hamiltonian is softened by a free-space SRG evolution at the three-body level to $\SRGpara=\SI{0.08}{\femto\meter^4}$  \cite{PhysRevLett.107.072501,PhysRevC.75.061001,PhysRevLett.103.082501,PhysRevC.90.024325,PhysRevC.85.021002,Wendt:2013bla}.
Details on the SRG evolution and the treatment of the 3N contributions can be found in Ref.\ \cite{PhysRevC.90.024325}.
All IM-SRG calculations are performed with a single-particle basis including 13 major shells ($e_{\max}=12$) and we have confirmed in all cases that the energies are sufficiently converged.
We truncate the initial three-body matrix elements in the total HO energy quantum number $E_{3\text{max}}=14$ in order to control the memory requirements \cite{Binder2014119,PhysRevC.87.034307,PhysRevLett.110.242501,PhysRevC.87.021303,Roth:2011vt}.
We employ a Hartree-Fock (HF) basis which largely eliminates the dependence of the IM-SRG on the oscillator frequency \cite{Hergert2016165}.

We obtain a reference state for a given nucleus from an initial NCSM calculation with the full NN+3N Hamiltonian for $\Nmaxref=0$.
The initial Hamiltonian is normal ordered with respect to this multi-determinantal reference state and the residual normal-ordered 3N piece is discarded. We have demonstrated that the effect of this truncation on ground-state energies is on the order of 1\% for the relevant mass range \cite{Roth:2011vt,PhysRevC.93.031301}.
For reasons of efficiency, we use a $J$-coupled formulation of the multi-reference IM-SRG that requires a reference state with vanishing total angular momentum, which limits the present discussion to even particle numbers.
However, the theoretical framework is completely general and one would only need to implement the IM-SRG for non-scalar tensor operators.
Alternatively, we can use the same ideas as in particle-attached or particle-removed equations-of-motion CC calculations to tackle odd nuclei \cite{PhysRevC.83.054306,QUA:QUA22367}.

\paragraph{Ground-State Energies.}

\begin{figure}
\includegraphics[width=\columnwidth]{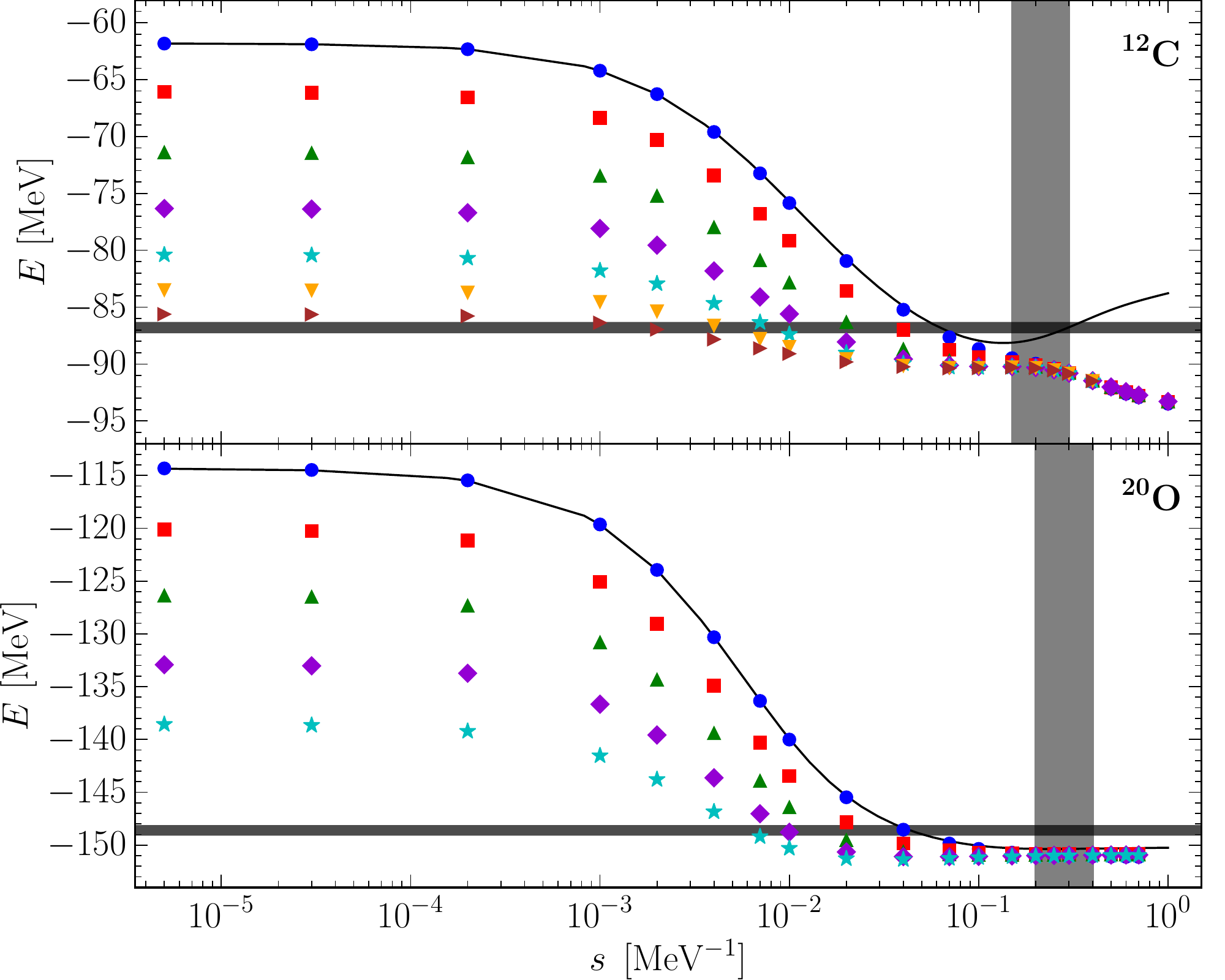}
\caption{(color online) Evolution of the ground-state energies in \elem{C}{12} and \elem{O}{20}.
Depicted is the zero-body part of the flowing Hamiltonian $\MEzeroHamNO(\IMSRGpara)$ (black solid line) and the lowest eigenvalue of $H(\IMSRGpara)$ obtained in IM-NCSM calculations for \Nmax=0~(\NmaxZeroSymbol), 2~(\NmaxTwoSymbol), 4~(\NmaxFourSymbol), 6~(\NmaxSixSymbol), 8~(\NmaxEightSymbol), 10~(\NmaxTenSymbol) and 12~(\NmaxTwelveSymbol).
All calculations use a HF basis with $\emax=12$ and $\hbar\Omega=\SI{20}{\MeV}$.
The horizontal grey band shows the importance-truncated NCSM result with explicit 3N interaction in the HO basis including the extrapolation uncertainties.
The vertical gray band represents the range of flow parameters $\IMSRGpara_{\max}/2$ to $\IMSRGpara_{\max}$ for the quantification of uncertainties (see text).
}
\label{fig:flowEgs}
\end{figure}

We start by investigating the evolution of ground-state energies of selected open-shell nuclei.
Figure~\ref{fig:flowEgs} shows the ground-state energies for \elem{C}{12} and \elem{O}{20} obtained from IM-NCSM calculations for several values of $\Nmax$ as a function of the flow parameter $\IMSRGpara$.
For comparison we also show the zero-body piece of the flowing Hamiltonian, i.e., the expectation value $E(\IMSRGpara) = \refstatebra  H(\IMSRGpara) \refstateket$.

A first important observation concerns the convergence of the final NCSM calculations with $\Nmax$.
For small flow parameters we recover the results of a direct NCSM calculation with the initial normal-ordered Hamiltonian in the HF basis, which exhibits very slow convergence. With increasing IM-SRG flow parameter the NCSM convergence is accelerated up to a point where all model spaces, including $\Nmax=0$, yield practically the same energy eigenvalue. This is proof that the IM-SRG successfully decouples the reference state and the $\Nmax=0$ space from all basis states at higher $\Nmax$.

A second observation pertains to the role of the zero-body part of the flowing Hamiltonian.
In the initial stages of the evolution the expectation value $E(\IMSRGpara)$ and the lowest eigenvalue  agree, i.e., the reference state remains an $\Nmax=0$ eigenstate of $H(\IMSRGpara)$ to a good approximation. 
However, in some cases, the $\Nmax=0$ eigenvalue of the evolved Hamiltonian is below $E(\IMSRGpara)$ in the later stages of the flow, i.e., the reference state is not an $\Nmax=0$ eigenstate anymore.
This is not surprising, since the IM-SRG transformation changes the structure of the Hamiltonian within the $\Nmax=0$ space. Therefore, $E(\IMSRGpara)$ loses its interpretation as ground-state energy and we have to explicitly diagonalize $H(\IMSRGpara)$. 
The effect is much stronger for the ground-state energy of \elem{C}{12} than for \elem{O}{20}.

A third observation concerns the many-body contributions that are discarded due to the truncation of the IM-SRG flow equations at the normal-ordered two-body level. Their effect can be estimated by comparing with importance-truncated NCSM results in the HO basis that include explicit 3N interactions. For \elem{O}{20} we find a deviation of less than $2\%$, which is in line with previous IM-SRG calculations \cite{PhysRevLett.110.242501}. Earlier studies have shown that the truncation of the initial normal-ordered Hamiltonian alone causes a deviation of approximately $1\%$ \cite{Roth:2011vt,PhysRevC.93.031301}. For \elem{C}{12}, which is a special case, the deviations are larger, slightly above $4\%$, and we observe a distinctive drop after a plateau of stable well-converged energies, signalling a systematic growth of induced many-body contributions.

To extract the final ground-state energy, we select a maximum flow parameter $\IMSRGpara_{\max}$ within the plateau for which stable convergence is observed at sufficiently small $\Nmax$.
Additionally, we consider the energy much earlier in the evolution, i.e., at $\IMSRGpara_{\max}/2$.
The range from $\IMSRGpara_{\max}/2$ to $\IMSRGpara_{\max}$ (cf. Fig.~\ref{fig:flowEgs}) provides an uncertainty estimate for the energy at a given $\Nmax$.
If the  evolution is stable and saturates, this uncertainty is very small.
Only if the evolution fails to stabilize or the $\Nmax$ convergence is incomplete, the uncertainty will be non-negligible.
We will use this uncertainty quantification protocol for all observables.

\begin{figure}
\includegraphics[width=\columnwidth]{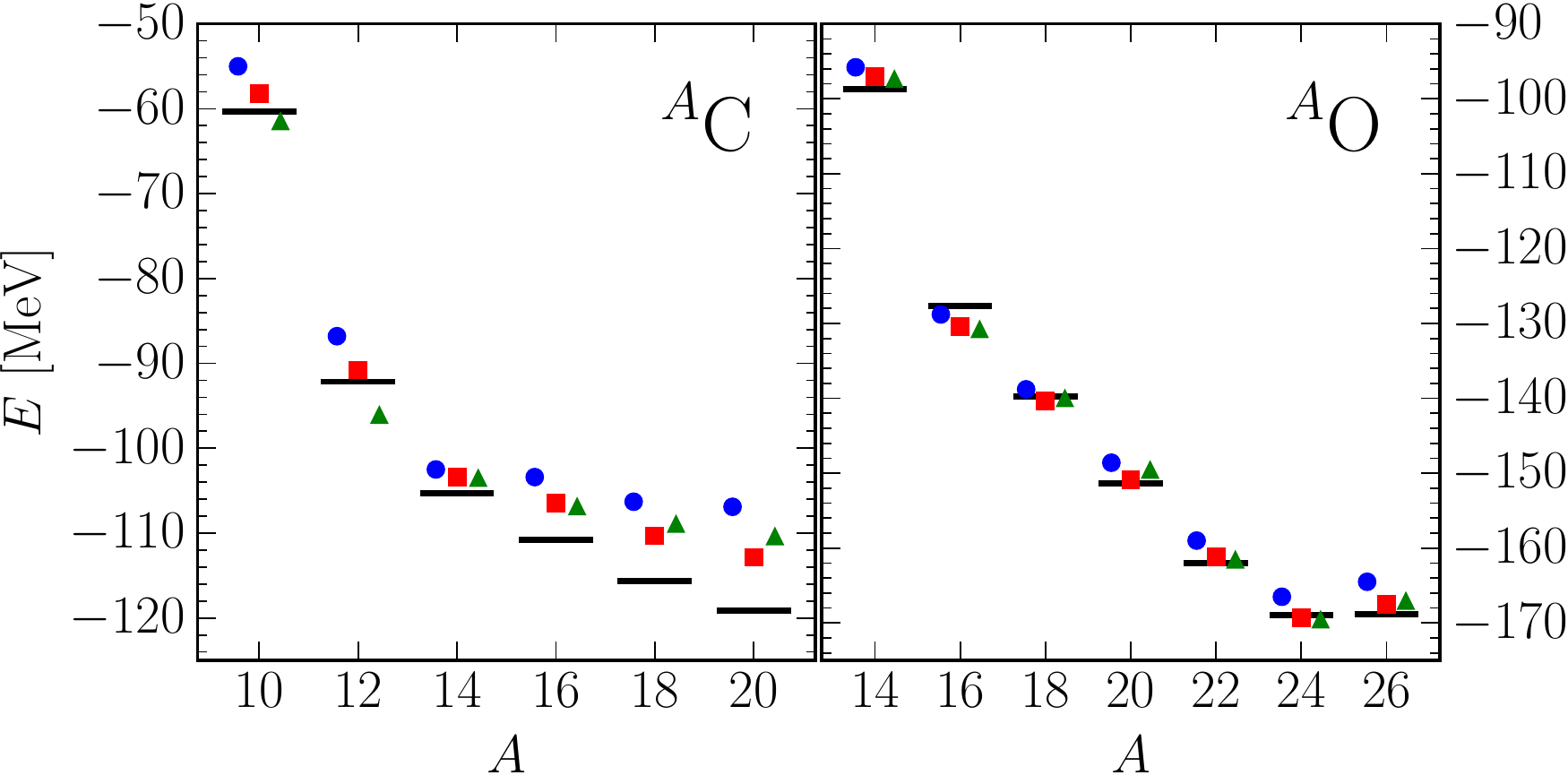}
\caption{(color online) Ground-state energies for even carbon and oxygen isotopes obtained from the IM-NCSM at $\Nmax=4$ (\IMNCSMsymbol) in comparison to importance-truncated NCSM calculations with explict 3N interactions (\ITsymbol) and the MR-IM-SRG with HF-Bogoliubov reference states (\MRIMSRGHFBsymbol) \cite{PhysRevLett.110.242501}.
Experimental values are indicated by black bars \cite{kaeri}.
The uncertainty due to flow-parameter dependence is negligible on this scale.
}
\label{fig:COchains}
\end{figure}

The IM-NCSM ground-state energies obtained for the even carbon and oxygen isotopes are summarized in Fig.~\ref{fig:COchains} and compared to results from importance-truncated NCSM calculations up to $\Nmax=12$ including all 3N contributions. The latter use a simple exponential extrapolation with an uncertainty of up to 1 MeV for the most neutron-rich isotopes. We also show multi-reference IM-SRG results obtained with spherical number-projected HF-Bogoliubov reference states \cite{PhysRevLett.110.242501}. For the oxygen isotopes the results of all three methods agree very well, with maximum deviations between IM-NCSM and NCSM around $1.8\%$ for the heaviest isotopes. 
For the carbon isotopes we observe larger deviations among the three methods, with \elem{C}{12} being the most significant case. Already IM-NCSM and MR-IM-SRG with HF-Bogoliubov reference states differ by almost $6\%$ for \elem{C}{12}, while the methods agree very well for \elem{C}{14}. Similar deviations have been observed with valence-space interactions obtained from the IM-SRG \cite{Stroberg:2016ung} and more detailed investigations of this special case are in progress.


\paragraph{\color{black}Excitation Energies.}

\begin{figure}
\includegraphics[width=\columnwidth]{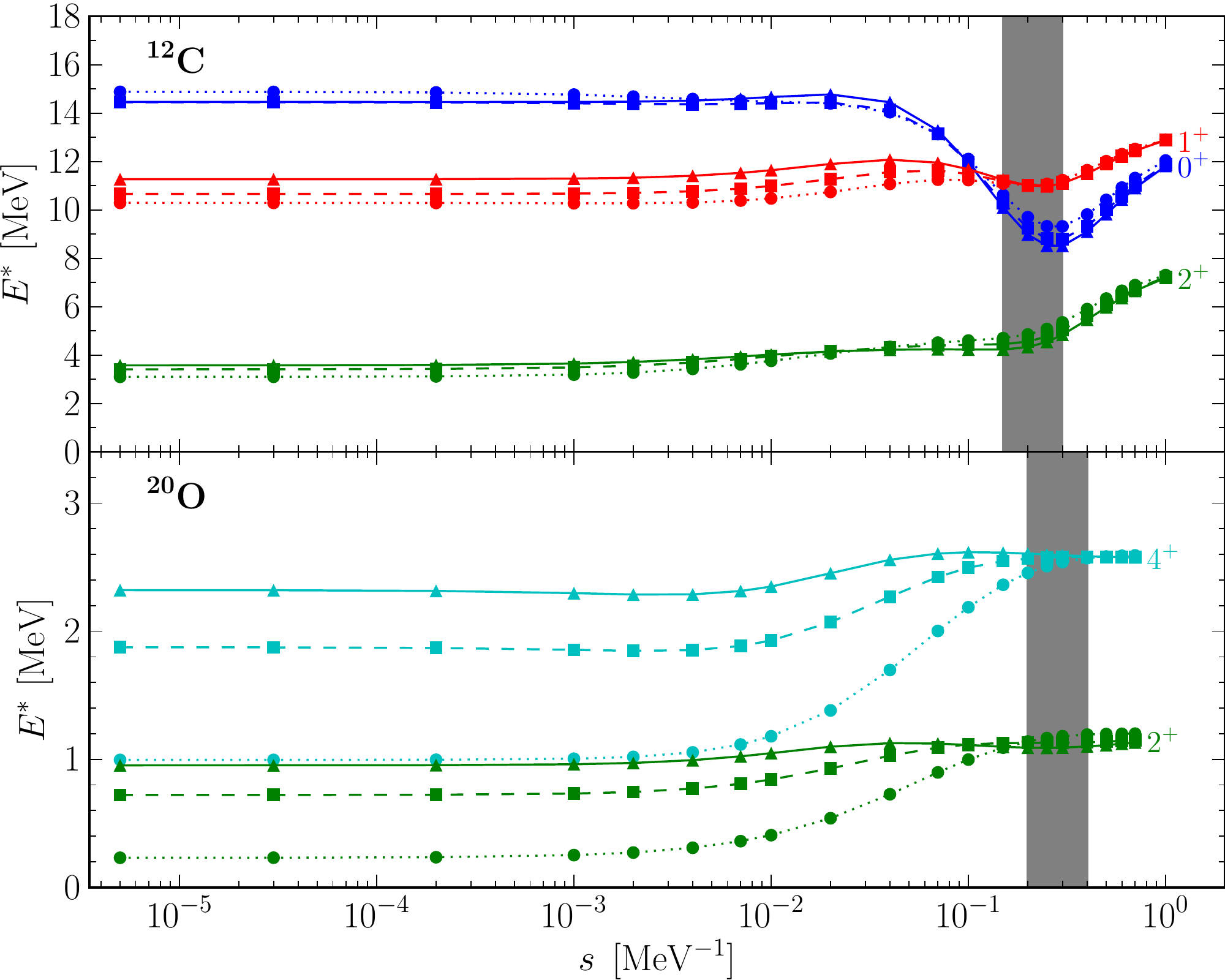}
\caption{(color online) Evolution of the excitation energies in \elem{C}{12} and \elem{O}{20}.
Depicted are the low-lying eigenvalues of $H(\IMSRGpara)$ obtained in NCSM calculations for $\Nmax=0$~(circle, dotted line), $2$~(square, dashed line), $4$~(triangle, solid line).
All calculations use a HF basis with $\emax=12$ and $\hbar\Omega=\SI{20}{\MeV}$.
The gray band represents the range of flow parameters for the uncertainty analysis.
}
\label{fig:flowEx}
\end{figure}

The IM-NCSM calculations automatically yield excited states and excitation energies, and we can analyze their flow-parameter dependence and convergence with $\Nmax$.
Figure~\ref{fig:flowEx} presents the evolution of the excitation energies for the first few excited states in \elem{C}{12} and \elem{O}{20} for $\Nmax=0,2,4$.
We add a HO Hamiltonian for the center of mass, evolved consistently in the IM-SRG, to remove spurious center-of-mass excitations from the spectrum \cite{GLOECKNER1974313}.
We observe that the rate of convergence of the excitation energies is not always improved.
However, the decoupling of the reference state from all excitations causes the excitation energies to converge monotonically from above for sufficiently large flow parameters.

For \elem{O}{20} the excitation energies become independent of the flow parameter and exhibit perfect convergence in the same region as the ground-state energy (gray band). For \elem{C}{12} the excitation energies also stabilize in that region, but then start to increase as a consequence of the distinctive drop of the ground-state energy discussed in Fig.~\ref{fig:flowEgs}.
Up to the flow parameter $\IMSRGpara_{\max}$ determined from the ground-state evolution, the dependence of the excitation energies is weak once we reach $\Nmax=4$.

The $0^+$ excited state in \elem{C}{12} is a notable exception.
In the region of flow parameters where the decoupling of the ground state occurs, the excitation energy of this $0^+$ state drops from about $\SI{14}{\MeV}$ to $\SI{8}{\MeV}$, while the other excitation energies remain stable. We have confirmed that this effect is robust under variation of the single-particle basis, the IM-SRG generator, and the center-of-mass constraint.
In conventional NCSM calculations the $0^+$ state at $14$ MeV is well known \cite{PhysRevC.90.014314} and believed to represent the Hoyle state, a three-alpha cluster state that cannot be converged in standard NCSM calculations \cite{PhysRevLett.98.032501,1742-6596-403-1-012028}.
The IM-SRG evolution seems to decouple multi-particle multi-hole excitations---needed to describe the Hoyle state---from the reference space, so that the $\Nmax=0$ result is already much better than the largest possible direct NCSM calculations. We also evaluated mass and charge radii and the electric monopole transition matrix-element for the first excited $0^+$ state. They increase rapidly as the excitation energy decreases, approaching the large values characteristic for the Hoyle state  \cite{PhysRevLett.98.032501,PhysRevLett.109.252501,PhysRevC.80.054603}. Unfortunately, before the excited $0^+$ state stabilizes, the discarded many-body terms in the IM-SRG flow equations start to distort the results (beyond the gray band). In order to quantitatively explore the Hoyle state, we are working on the inclusion of these contributions.

\begin{figure}
\includegraphics[width=\columnwidth]{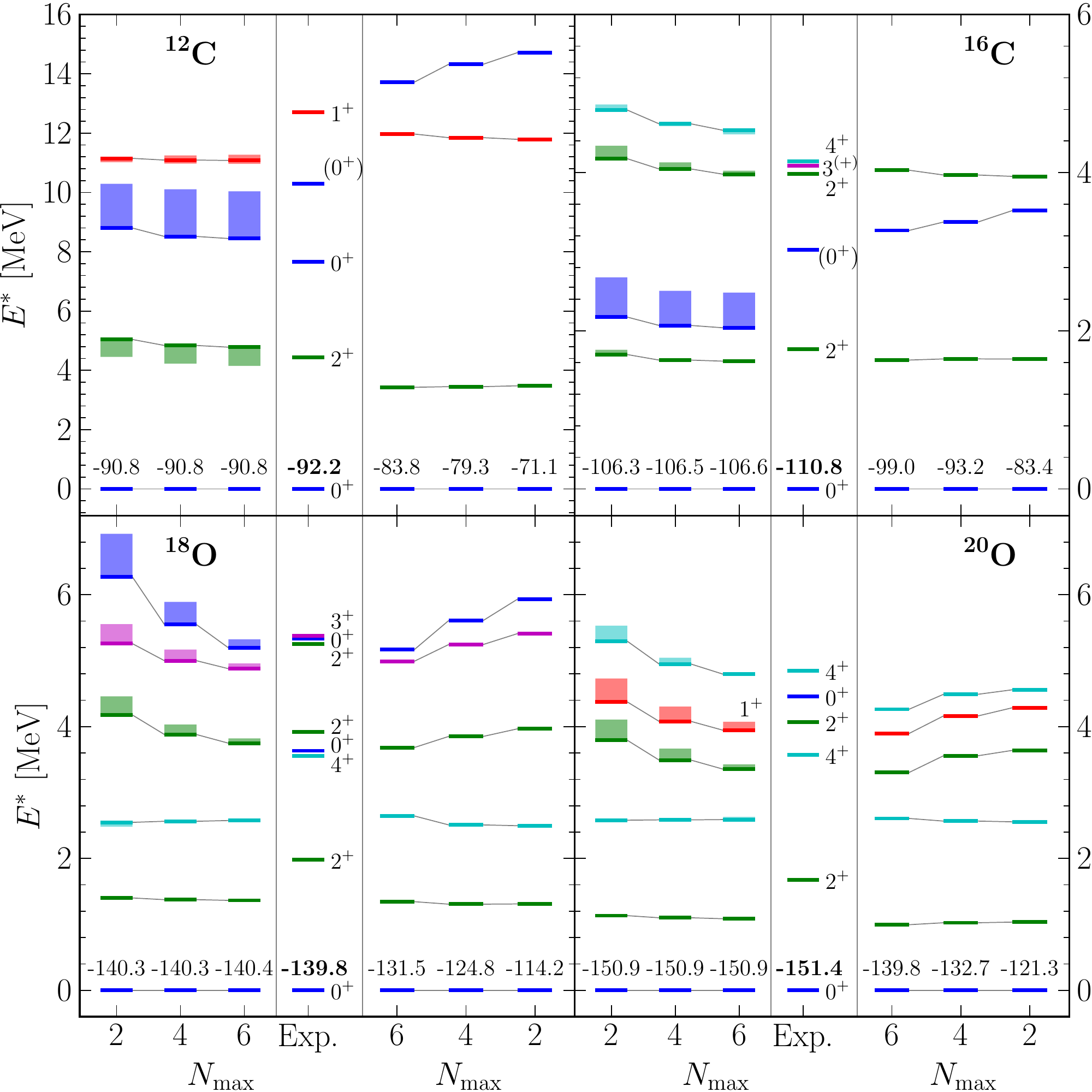} 
\caption{(color online) Excitation spectra for selected isotopes obtained from the IM-NCSM  (left-hand columns) and the importance-truncated NCSM (right-hand columns) in comparison to experiment (center column) \cite{nndc}.
The bands indicate the residual flow-parameter dependence in the range from $\IMSRGpara_{\max}/2$ to $\IMSRGpara_{\max}$ with bars marking the latter.
The importance-truncated NCSM results are obtained with explicit 3N interactions using the HO basis with $\hbarOmega=\SI{16}{\MeV}$.
}
\label{fig:COspectra}
\end{figure}

The excitation spectra of selected nuclei and their $\Nmax$ convergence obtained in IM-NCSM and importance-truncated NCSM calculations are presented in Fig.~\ref{fig:COspectra}.
The bands for the IM-NCSM result from the uncertainty estimate described earlier, using the same range of $s$ as for the ground-state energies.
We find excellent agreement for the excitation energies of states that are robust and well converged, indicating that all truncation effects are small.
As for \elem{C}{12}, we find an excited $0^+$ state in the IM-NCSM for \elem{C}{16} which lies at higher energy in the conventional NCSM.
Some higher lying states, however, show a slower convergence indicating an intrinsic structure that probes pieces to the Hamiltonian that are not completely decoupled. This is an interesting aspect for further optimizations of the IM-SRG generators.

\paragraph{Conclusions.}

We have merged the NCSM with the multi-reference IM-SRG to define a new \emph{ab initio} many-body method capable of describing ground and excited states of closed and open-shell nuclei on the same footing.
In simple terms, the approach can be viewed as NCSM calculation with an in-medium decoupled Hamiltonian and combines the advantages of the NCSM and the IM-SRG.
We have demonstrated its efficiency and accuracy in comparison to direct NCSM calculations, which are computationally much more demanding. In contrast to valence-space shell-model calculations based on effective interactions derived from IM-SRG \cite{PhysRevLett.113.142501} or CC \cite{PhysRevLett.113.142502}, the IM-NCSM is a no-core approach and convergence with respect to all model-space truncations can be demonstrated explicitly.
For benchmark purposes, we limited ourselves to systems that are accessible to the conventional NCSM, but the moderate computational scaling will allow us to study the full range of medium-mass nuclei.
In future work, we will demonstrate the extension to odd systems and electromagnetic observables and investigate the specific aspects of \elem{C}{12} in more detail.

\begin{acknowledgments}
\paragraph{Acknowledgments.} 

This work is supported by the Deutsche Forschungsgemeinschaft through grant SFB 1245, the Helmholtz International Center for FAIR, and the BMBF through contracts 05P15RDFN1 (NuSTAR.DA) and 05P2015 (NuSTAR R\&D). Heiko Hergert acknowledges startup support from the NSCL/FRIB Laboratory and Michigan State University.
The authors gratefully acknowledge the computing time granted on the supercomputers JURECA at the J\"ulich Supercomputing Centre, LOEWE at the Center for Scientific Computing Frankfurt, and Lichtenberg at the Technische Universit\"at Darmstadt. 
Further computing resources were provided by the Michigan State University High Performance Computing Center (HPCC)/Institute for Cyber-Enabled Research (iCER), and the National Energy Research Scientific Computing Center (NERSC), a DOE Office of Science User Facility supported by the Office of Science of the U.S.~Department of Energy under Contract No. DE-AC02-05CH11231.

\end{acknowledgments}


%

\end{document}